$NaCaNi_2F_7$: A frustrated high temperature pyrochlore antiferromagnet with S=1 $Ni^{2+}$

J. W. Krizan[1*] and R. J. Cava[1]

[1]Department of Chemistry, Princeton University, Princeton, NJ 08544, USA

* Corresponding Author: jkrizan@princeton.edu

Abstract:

$NaCaNi_2F_7$ is an insulating, frustrated $A_2B_2F_7$ pyrochlore with magnetic S = 1 $Ni^{2+}$ on the pyrochlore B site. Non-magnetic Na and Ca are disordered on the A-site. Magnetic susceptibility measurements made on an oriented single crystal, grown in a floating zone furnace, show isotropic behavior at temperatures between 5 and 300 K, with an effective moment of 3.7 $\mu_B$/Ni. Despite displaying a large Curie-Weiss theta (-129 K), spin-ordering-related features are not seen in the susceptibility or specific heat until a spin glass transition at 3.6 K. This yields an empirical frustration index of $f$ = -$\theta_{CW}$/$T_f$ ≈ 36. The spin glass behavior is substantiated by a shift of the freezing temperature with frequency in the AC susceptibility, bifurcation in the DC susceptibility, and by a broad maximum in the magnetic specific heat. The observations as made on large single crystals suggest that $NaCaNi_2F_7$ is likely a realization of a frustrated spin 1 pyrochlore antiferromagnet with weak bond disorder.

**Introduction:**

The lattice of corner-sharing metal atom tetrahedra in pyrochlores is one of the well-known magnetic motifs where nearest neighbor antiferromagnetic interactions cannot be strictly satisfied, and as a result, rare earth (R) pyrochlores have been intensely studied as model geometrically frustrated magnets. Research has focused on those with Ti and Sn on the non-magnetic sites, and especially on the $R_2Ti_2O_7$ pyrochlores due to the availability of large single crystals grown by the floating zone method. These oxide pyrochlores host different low temperature phenomena such as long range ordered, spin ice, spin glass, and spin liquid ground states[1,2]. The low magnetic coupling strengths in rare earth pyrochlores add complexity to the modeling and measurements, as small perturbations in structural characteristics can often have significant effects on the magnetic ground states, which manifest at low temperatures. Using the $R_2Ti_2O_7$ pyrochlores as inspiration, we seek analogous systems based on magnetic transition elements due to their potentially larger magnetic interactions. This has motivated us to resurrect the little known fluoride pyrochlores of the $A^+A'^{2+}B_2^{2-}F_7$ type; very few fully fluorine-based pyrochlores are known[3,4], and fewer still have been grown as single crystals and characterized[4,5].

The experimentally determined frustration index (*f*) is expressed as f = -$\theta_{CW}$/T$_f$, where $\theta_{CW}$ is the Weiss temperature and T$_f$ is the temperature where an ordering-related feature is observed by susceptibility or specific heat. The frustration index is frequently employed as an empirical one-parameter indication of the degree of magnetic frustration in a system.[6] Here we report our characterization of cm-scale $NaCaNi_2F_7$ single crystals, which show a large moment of 3.7 $\mu_B$/Ni, a high Curie-Weiss theta (-129 K), and no spin freezing until 3.6 K, yielding a frustration index of 36. While the moment is large, it is still well below what would be possible if the full orbital contribution for $Ni^{2+}$ is present (5.59 $\mu_B$/Ni). Specific heat measurements show that a substantial amount of the spin entropy is released at the 3.6 K transition, indicating a large majority or all of the spins participate in the spin glass ordering. Structurally, Na and Ca are found to be fully disordered on the A site of the pyrochlore with the Ni and F sublattices fully ordered. The magnetic properties observed for $NaCaNi_2F_7$ suggest that the A-site disorder likely induces weak bond disorder in the Ni-Ni interactions. This type of bond disorder has been described theoretically, and is expected to precipitate a spin glass state under the circumstances found in $NaCaNi_2F_7$.[7–9] Bond disorder has similarly been proposed to explain the enigmatic properties of the nominally structurally ordered 4*d* metal oxide pyrochlore $Y_2Mo_2O_7$[10]. In the fluoride systems, the origin of the bond disorder (the Ca/Na mixing) appears to be far clearer.

In many other materials commonly referred to as pyrochlores, the "pyrochlore network" designation refers to the arrangement of the magnetic cations only, with the remainder of the structure quite different from that of a traditional pyrochlore - thus these compounds are not ideally analogous to the $R_2Ti_2O_7$ materials or the fluorides studied here.[11,12] While single crystals are available of $Y_2Mo_2O_7$, other transition-metal-based traditional oxide pyrochlores, many of which show antiferromagnetic interactions and exhibit glassy behavior, are not available as large single crystals and do not show the same level of frustration as the fluorides.[10,12,13,14] There is theoretical interest in the $FeF_3$ pyrochlore, but unfortunately, the methods employed for its synthesis likely preclude the growth of this material as large single crystals. In contrast, the S = 1 $NaCaNi_2F_7$ and S = 3/2 $NaCaCo_2F_7$ and $NaSrCo_2F_7$ fluoride pyrochlores have now been grown as cm-scale single crystals, suggesting that transition metal fluorides have the potential develop into a significant and readily accessible new family of geometrically frustrated magnets.[4,5]

**Experimental**

Large single crystals of NaCaNi$_2$F$_7$ were grown by a modified Bridgman-Stockbarger method[15,16] in an optical floating zone furnace (see supplemental material.) Elemental fluorides were pre-reacted in a custom-built alloy 400 reactor in an atmosphere of anhydrous HF and argon. Throughout the synthetic procedure, air and moisture were rigorously excluded due to the hygroscopic nature of NiF$_2$. The structural characterization of NaCaNi$_2$F$_7$ was carried out at 100 K by single crystal X-ray diffraction (see supplemental material). Temperature dependent DC susceptibility measurements were carried out in a Superconducting Quantum Interference Device (SQUID) equipped Quantum Design Magnetic Property Measurement System (Model MPMS XL-5). Field dependent DC susceptibility measurements, AC susceptibility measurements, and heat capacity measurements were made in a Quantum Design Physical Property Measurement System (PPMS). All magnetization measurements used the same oriented single crystal. The heat capacity was extracted by the heat relaxation method on a 7.1 mg single crystal. The crystal platelet used was mounted with Apiezon-N grease on a nonmagnetic sapphire stage. Low temperature heat capacity measurements utilized the Quantum Design $^3$He insert. The heat capacity of nonmagnetic NaCaZn$_2$F$_7$[5] was subtracted from that of NaCaNi$_2$F$_7$ to estimate the magnetic heat capacity.

**Results and discussion**

An image of the single crystal that was sectioned and employed in this study is shown in Figure 1; its light green color is characteristic of insulating Ni$^{2+}$ compounds. Figure 1b shows the hk0 plane of the Fd-3m reciprocal lattice of a NaCaNi$_2$F$_7$ crystal obtained at 100 K, which is fully consistent with what is expected for the classical pyrochlore structure; the observed reflections indicate that there is no long-range ordering of Na and Ca on the A-site. The quantitative structure refinements indicated the stoichiometry was within error of the expected NaCaNi$_2$F$_7$ and no A-B site mixing (i.e. stuffing or anti-stuffing) was detectable. The only variable structural parameter in the pyrochlore structure describes the position of the framework fluorine ion. This parameter, $x$ = 0.3303(2) for NaCaNi$_2$F$_7$, reflects NiF$_6$ octahedra with 6 equal Ni-F bond lengths of 2.0 Å with F-Ni-F bond angles, 83.2° and 96.8°, that deviate somewhat from the ideal 90° value. In the lower panel of Figure 1, a powder-diffraction pattern from a ground single crystal shows a characteristic pyrochlore pattern, in excellent agreement with a calculated pattern from the structural model determined by single-crystal x-ray diffraction. Diffraction experiments on finely ground powders from the single crystals exposed to laboratory air for a few days (not shown) indicate changes in diffracted peak intensity, suggesting decomposition of the samples on exposure to moisture in the air. No degradation was detected in the single crystals over the course of months, however, likely due to their small surface to volume ratio. As a precaution, crystals were stored in a desiccator.

The temperature dependent bulk DC susceptibility of NaCaNi$_2$F$_7$ is presented in Figure 2. The applied field (H)-dependent magnetization (M) at 2 K (upper inset) shows a linear response up to $\mu_0$H = 9 T with no signs of saturation. The susceptibility was therefore defined as M/H for $\mu_0$H = 0.2 T. The lower inset shows the susceptibility data of a single crystal with the field applied in the [100], [110], and [111] directions. The observed susceptibilities are equal, within experimental error, as is expected for a second rank tensor property in this space group. For the sake of clarity, the inverse susceptibility of only the [100] direction is therefore given in the main panel. The Curie-Weiss law ($\chi=C/T-\theta$) is fit from 150 to 300 K. The fit shows a large effective moment, $p_{eff}$ = 3.7(1), for S=1 $Ni^{2+}$, indicating the presence of a significant, but not maximum, orbital contribution.[17] The Curie-Weiss temperature of -129 K indicates strong antiferromagnetic interactions, but no features are seen in the susceptibility until roughly 3.6 K. This yields a frustration index of 36.

By rearranging the Curie-Weiss law ($\chi=C/T-\theta$) to $C/(\chi|\theta|)=T/|\theta|-1$ it is possible to create a normalized, dimensionless plot that is useful for comparing the magnetic behavior of related materials.[18] In Figure 3, ideal antiferromagnets would follow the line y=x+1 (shown as a dashed line), with indications of magnetic ordering on the order of $T/|\theta| \approx 1$. Using this representation, the magnetic behavior of $NaCaNi_2F_7$ is compared to $NaCaCo_2F_7$.[5] The results of the Curie-Weiss fits are given in the upper left for reference. Looking at the middle of the main panel, it is evident both materials follow ideal Curie-Weiss behavior (dashed line) very closely at higher temperatures. $NaCaNi_2F_7$ may show a very slight ferromagnetic deviation from Curie Weiss behavior between 0.5 and 1.0 ($T/|\theta|$). As highlighted in the inset, both materials deviate from Curie-Weiss behavior at temperatures below 0.25 ($T/|\theta|$), but in opposite directions. This difference suggests at temperatures just above the spin glass ordering, the low temperature magnetic behavior for the S = 1 Ni pyrochlore is significantly different from that of the S = 3/2 $Co^{2+}$ pyrochlores, despite the other similarities of the Ni and the Co S = 3/2 fluoride pyrochlore systems.

To further characterize the behavior of $NaCaNi_2F_7$, it is helpful to look in more detail at the low-temperature susceptibility, shown in Figure 4. The upper panel shows the temperature-dependent AC susceptibility under an applied field of 20 Oe, and a variety of excitation frequencies (10 and 50 Hz omitted for clarity). No additional features were observed in the AC susceptibility between 5 and 300 K (not shown.) A systematic upward temperature shift in the spin freezing is seen with increasing frequency ($\omega$.) This, in conjunction with the DC susceptibility shown in the middle panel, is a strong indication of the presence of a glassy transition in the magnetic system.

The DC susceptibility shows significant bifurcation between the zero-field-cooled and field-cooled data with an applied field of 200 Oe. This 3.6 K maximum in the DC susceptibility is an estimate of the freezing temperature ($T_f$) and is used to calculate the frustration index mentioned above. From the AC susceptibility measurements, the expression $\frac{\Delta T_f}{T_f \Delta log\omega}$ is useful for characterizing the spin-glass. A value of 0.024 is obtained for $NaCaNi_2F_7$, which like the $Co^{2+}$ fluoride pyrochlores falls into the range expected for an insulating spin glass. The value for the current material is slightly lower than 0.029 ($NaCaCo_2F_7$) and 0.027 ($NaSrCo_2F_7$).[4,5] Despite the presence of different magnetic ions, all three are very close. The spin glass behavior can be further parameterized by fitting the shift in the AC susceptibility to the Volger-Fulcher law, $T_f = T_0 - \frac{E_a}{k_b}\frac{1}{\ln(\tau_0 f)}$, which relates the freezing temperature and frequency to the intrinsic relaxation time ($\tau_0$), the activation energy of the process ($E_a$), and the ideal glass temperature ($T_0$). The $NaCaNi_2F_7$ fit is compared to that of $NaCaCo_2F_7$ in the bottom panel of Figure 4.[5] $T_f$ on the y-axis is scaled by the freezing temperature found from the DC susceptibility in order to compare the materials on the same scale. Given the very small temperature shift and the limited number of data points, a meaningful fit to all three parameters could not be obtained for $NaCaNi_2F_7$. Thus, $\tau_0$ was set to a physically meaningful value of $1\times10^{-12}$ for direct comparison to $NaCaCo_2F_7$ and $NaSrCo_2F_7$.[4,5] Values of $1.4(2)\times10^{-3}$ eV and 3.0(1) K were found for $E_a$ and $T_0$, respectively. The activation energy is very close to the other reported fluoride pyrochlores and as expected, the ideal glass temperature is slightly lower than the temperature observed in the DC susceptibility. A visual comparison of the fits shows the similarity of the materials. $NaCaNi_2F_7$ has a slightly higher activation energy and on a relative scale of $T_0/T_N$, the ideal glass temperature is lower. More precise measurements and further characterization are required to more accurately understand these subtle differences.

The magnetic heat capacity of $NaCaNi_2F_7$ yields further insight. The inset of the top panel of Figure 5 shows the raw heat capacity of $NaCaNi_2F_7$ and the previously reported nonmagnetic analog $NaCaZn_2F_7$.[5] The nonmagnetic analog was scaled down by 10.5%, which is more than for the Co analogs, to approximate the phonon contribution of $NaCaNi_2F_7$. The raw heat capacity shows $NaCaNi_2F_7$ diverges significantly from the nonmagnetic analog near the spin freezing transition. Through subtraction, this difference translates to the magnetic heat capacity peak in the top panel of Figure 5. The figure shows that the heat capacities at the peaks in $NaCaCo_2F_7$ and $NaCaNi_2F_7$ are roughly equivalent, but the peak in the latter is significantly broader and extends to higher temperatures.[5] Both materials show a saturation of the magnetic heat capacity by roughly 40 K. Integrating $C_p/T$ yields an estimate of the magnetic entropy freezing out in the transition, (lower panel); the value for $NaCaNi_2F_7$ is very close to that expected for a two-state magnetic system. As can be inferred from the top panel, there is more entropy released in $NaCaNi_2F_7$ than in $NaCaCo_2F_7$ in this temperature range, which we believe is beyond the experimental error.[5] Further work, however, is necessary to determine this conclusively.

The shape of the heat capacity below the freezing transition can sometimes give information regarding the elementary spin excitations in the material. The inset of the lower panel shows the heat capacity below the spin glass transition is linear on a $C(T) \propto T^2$ scale. For a spin glass, it is expected the heat capacity is linear on a $C(T) \propto T$ scale, arising from a distribution of two level systems linear in energy.[6] Despite the other magnetic behavior observed for $NaCaNi_2F_7$, consistent with a classical spin glass, linear behavior of the heat capacity is not observed. The behavior observed, $C(T) \propto T^2$, is expected for spin waves in a quasi-2D antiferromagnet. Further research is required to better understand this behavior since the pyrochlore lattice is typically considered a 3D frustrated geometry. There is some curvature still present in $NaCaCo_2F_7$ on these scales so no conclusion was obtained regarding the excitations.[5]

**Conclusion**

Despite the Curie-Weiss fit suggesting the presence of strong antiferromagnetic interactions on the order of 130 K, spin freezing does not occur until 3.6 K in $NaCaNi_2F_7$. This difference indicates this material is a highly frustrated S = 1 $Ni^{2+}$ pyrochlore. The crystal structure determination shows that Na and Ca are completely disordered over the A-site, and the unit cell dimension is in good agreement with the previous report.[3] Disorder results in a random local environment around the B-site magnetic cation, as has been seen in other pyrochlores.[5,19,20] Theoretical studies of the Heisenberg antiferromagnet on the pyrochlore lattice indicate that weak randomness in the exchange interactions (i.e. magnetic bond disorder) can precipitate a spin glass ground state, with the spin glass temperature set by the strength of the bond disorder.[7–9] We propose that that is the case here. The robustness and similarity seen in the transition metal fluoride pyrochlores offers an opportunity to explore the structure property relationships in a comparatively unexplored class of materials. The similarity in magnetic behavior of the $Co^{2+}$ ($3d^7$) and $Ni^{2+}$ ($3d^8$) -based fluoride pyrochlores is surprising, as these two ions do not typically behave similarly in solids. Further experiments to probe the magnetic properties are necessary and will be facilitated by the availability of large single crystals and readily accessible temperature range.

**Acknowledgements**

This research was conducted under the auspices of the Institute for Quantum Matter at Johns Hopkins University, and supported by the U. S. Department of Energy, Division of Basic Energy Sciences, Grant DE-FG02-08ER46544.

Figure Captions:

**Figure 1:** Top left panel: the oriented single-crystal boule of NaCaNi$_2$F$_7$ used in this study, with the indexed faces labeled. Top middle panel: the single-crystal precession image of the (0kl) reciprocal lattice plane for NaCaNi$_2$F$_7$ at 100 K. No superlattice reflections are seen, indicating the absence of Na-Ca long-range ordering. Top right panel: the NiF$_6$ octahedra in NaCaNi$_2$F$_7$ are slightly distorted along <111> with equivalent bond lengths of 2.0 Å. but F-Ni-F bond angles of 83.2° and 96.8°. Bottom panel: the powder-diffraction data showing that the bulk material of the crystal boule is an A$_2$B$_2$F$_7$ pyrochlore. The pattern fits the calculated pattern from the structure model determined by single crystal diffraction. The unit cell is 10.3027(2) Å at 300 K. Green tics show the expected peak positions.

**Figure 2:** Top inset: the field-dependent DC susceptibility at 2 K with the field applied along the [100] crystallographic direction. Bottom inset: temperature-dependent DC susceptibility with a field of 2000 Oe applied parallel to the [100], [110], and [111] crystallographic directions indicating isotropic behavior. Main panel: the inverse susceptibility and Curie-Weiss fit to the [100] data. Despite the large θ$_{CW}$ = -129 K, no spin freezing is seen until 3.6 K.

**Figure 3:** A dimensionless, normalized plot of the magnetization to compare the behavior of NaCaNi$_2$F$_7$ to NaCaCo$_2$F$_7$,[5] based on the normalized Curie-Weiss behavior, C/(χ|θ|)=T/|θ|-1. The main panel shows both materials follow Curie Weiss behavior to low temperatures. The inset highlights the low T/|θ| region near the transitions. Here NaCaNi$_2$F$_7$ shows an antiferromagnetic deviation from ideal behavior and NaCaCo$_2$F$_7$ shows a smaller ferromagnetic deviation from ideal behavior.

**Figure 4:** Top panel: the frequency dependence of the AC susceptibility in an applied field of 20 Oe. Middle panel: bifurcation in the DC susceptibility with an applied field of 200 Oe between the zero-field-cooled and field-cooled data. Bottom panel: parameterization of the shift in the AC susceptibility by fitting the frequency-dependent freezing temperature to the Volger-Fulcher law. The fit is also shown for NaCaCo$_2$F$_7$.[5] To aid in comparison, the y-axis is scaled by the freezing temperature, as determined by the maximum in the DC susceptibility.

**Figure 5:** The heat capacity, scaled per mol of the B-site cation (M). The inset to the top panel shows the raw heat capacity of NaCaNi$_2$F$_7$ and the nonmagnetic analog NaCaZn$_2$F$_7$[5] (scaled, used for subtraction to yield the magnetic heat capacity). The middle panel compares NaCaNi$_2$F$_7$ to NaCaCo$_2$F$_7$.[5] In the bottom panel, the entropy from the magnetic heat capacity is compared to the Ising (R ln(2)) and Heisenberg (R ln(2S+1)) limits. Inset, the temperature dependence of the heat capacity below the spin-freezing transition shows roughly C(T)∝T$^2$ behavior for NaCaNi$_2$F$_7$ and curvature in the heat capacity of NaCaCo$_2$F$_7$.

Figures:

Figure 1:

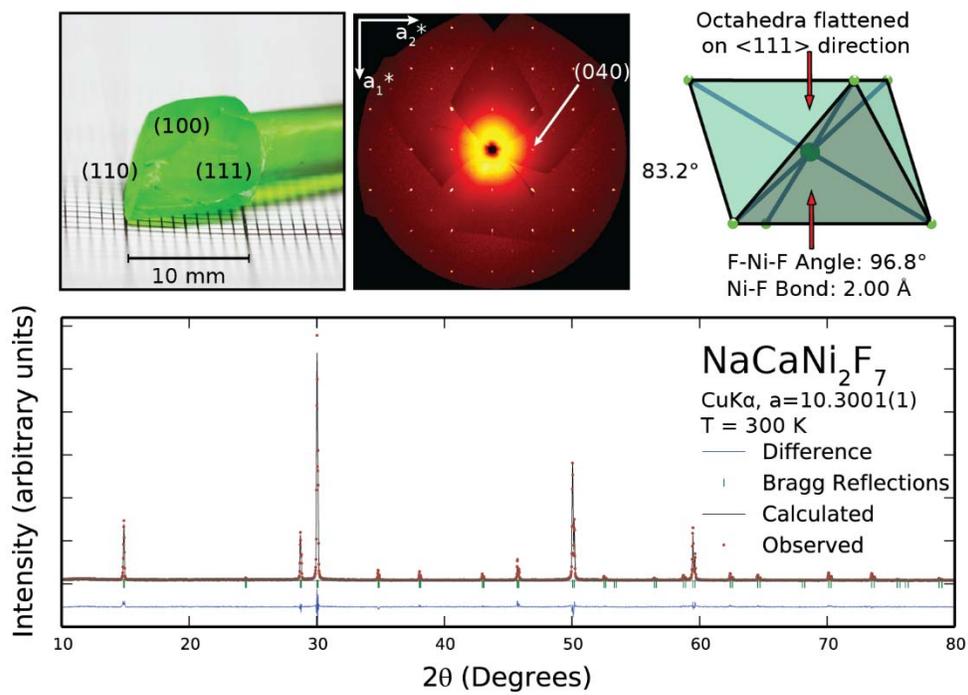

Figure 2:

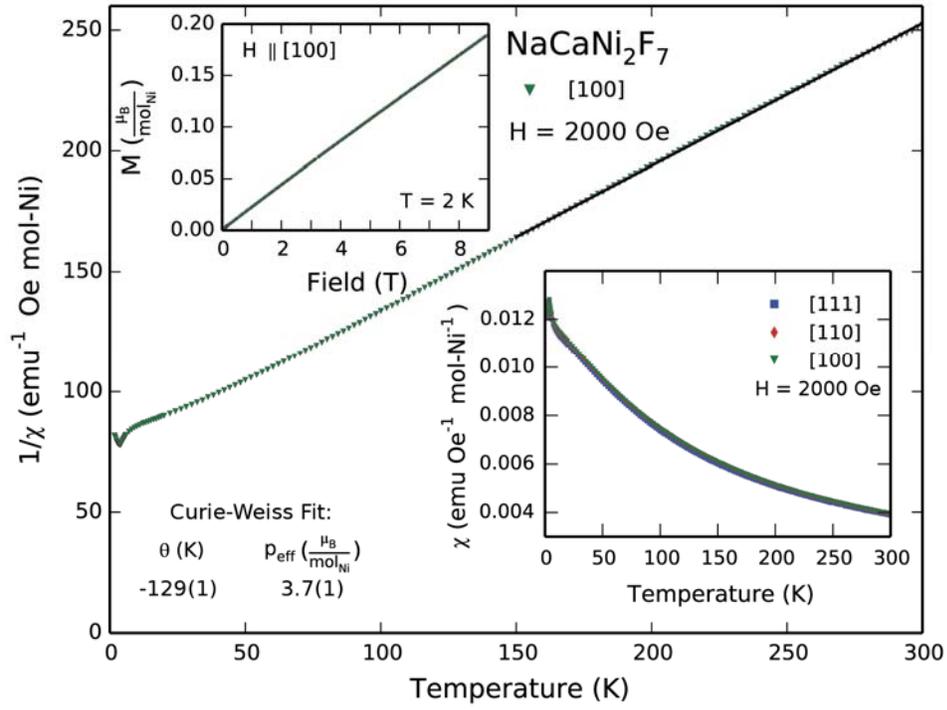

Figure 3:

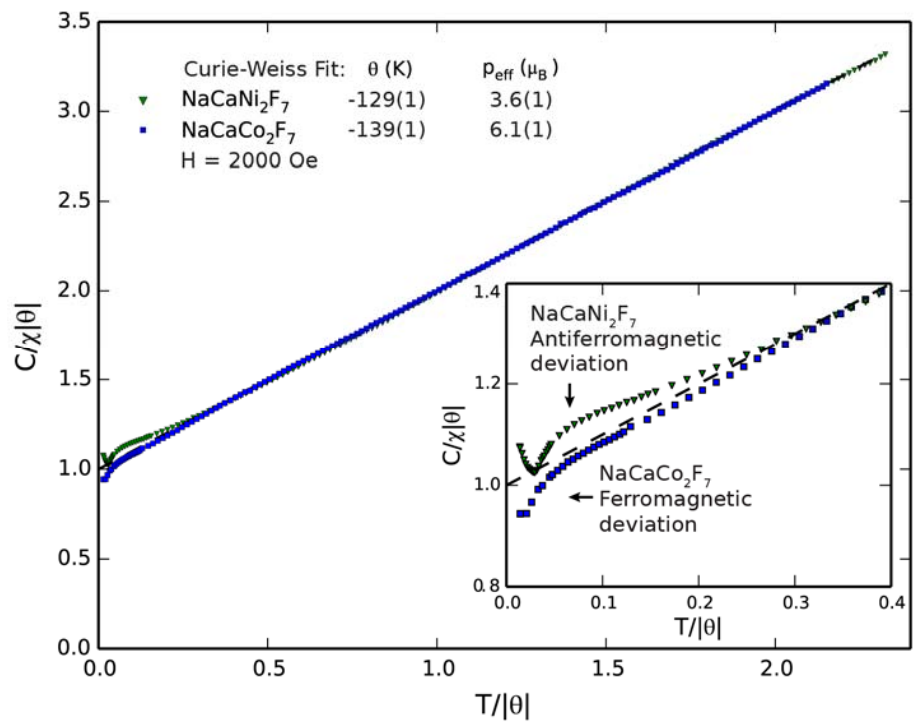

Figure 4:

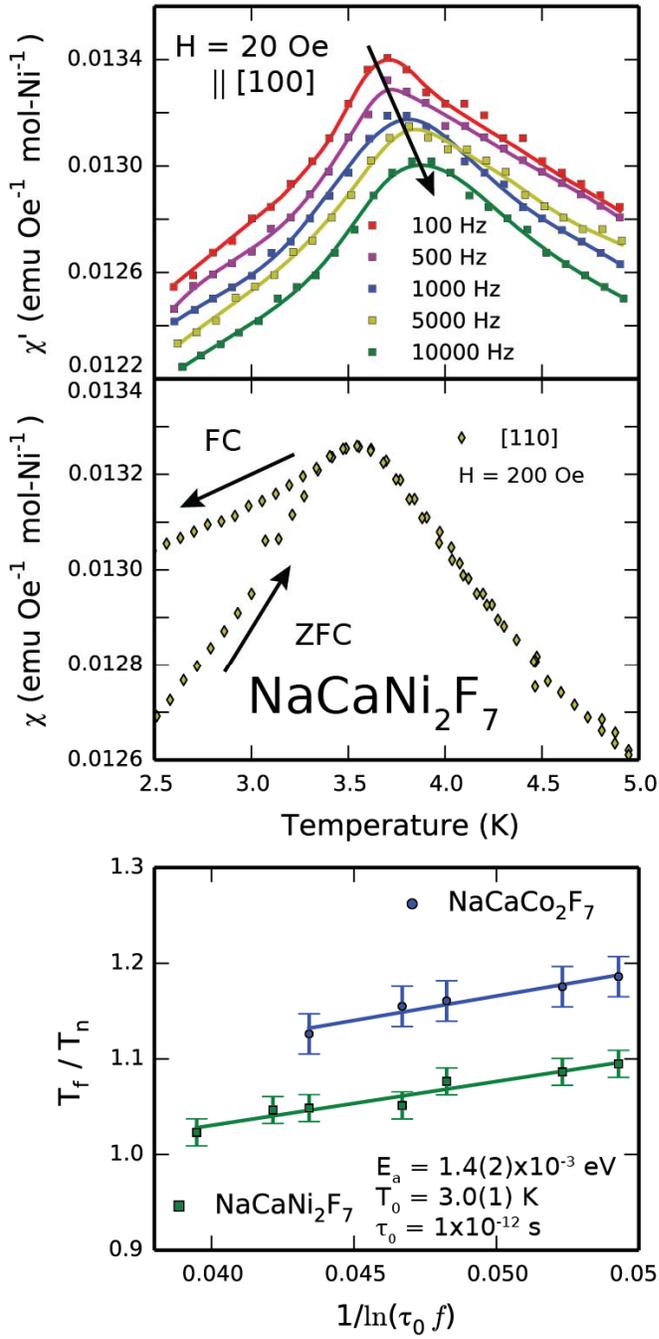

Figure 5:

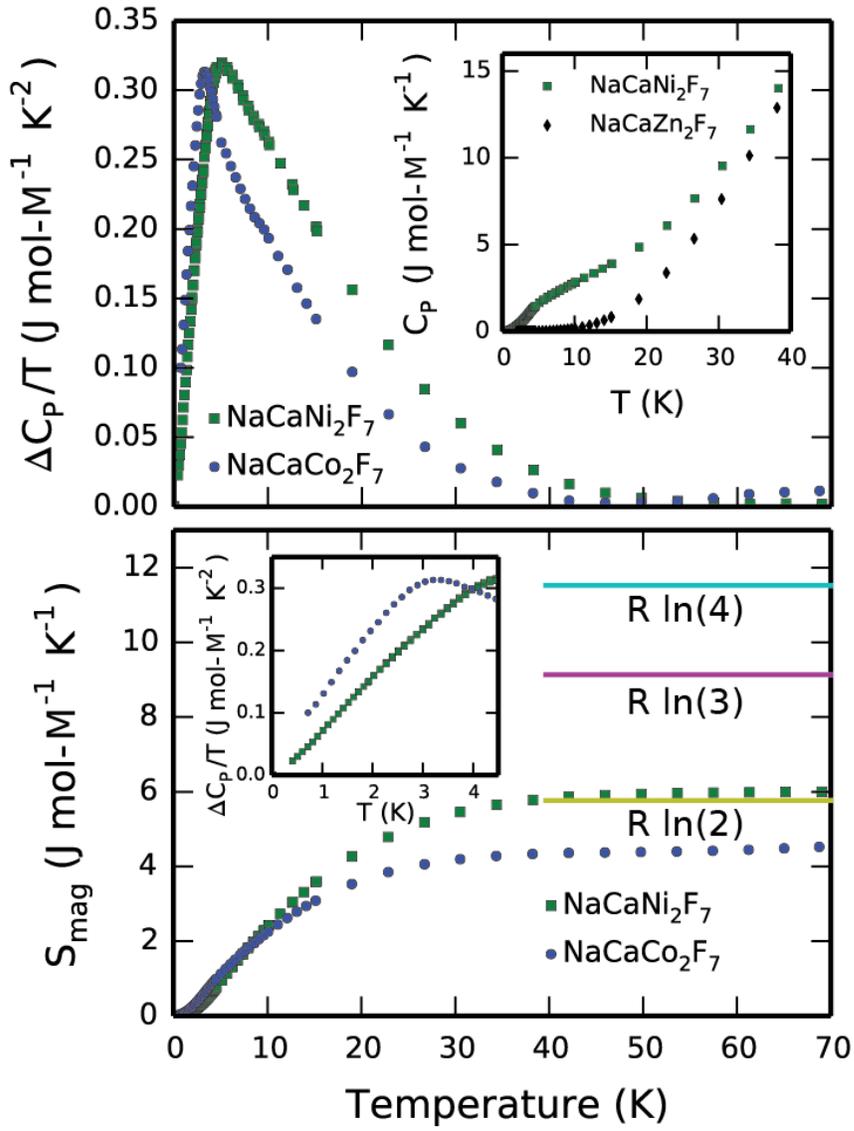

Supplemental Material:

## Supplemental Material

### Crystal Growth

All work with these materials was conducted in an argon glove box. Before beginning, NaF and $CaF_2$ were dried under dynamic vacuum at 175 °C for 24 hours. $NiF_2$ was heated at 600 °C under dynamic vacuum for eight hours in a custom-made alloy 400 reactor. The $NiF_2$ was heated subsequently under an anhydrous HF/argon atmosphere at 700 °C for 24 hours. The elemental fluorides were combined with a 10% excess of $NiF_2$ and ground thoroughly. These were then pre-reacted under an anhydrous HF/argon atmosphere at 700 °C for 35 hours.

The pre-reacted powder was loaded into a custom-made graphite mold and transferred to the optical floating-zone furnace for synthesis and crystal growth by the modified Bridgeman-Stockbarger method. The furnace used was a Crystal Systems Corporation FZ-T-10000-H-VI-VPO, equipped with 500 W halogen bulbs. A crystal growth rate of 3.5 mm/hr under a dynamic atmosphere of 8.5 bar Ar was successfully achieved. Some nickel metal appeared on the top surface of the as grown boule. This was removed mechanically and no inclusions were visible. As a precaution, a significant amount of extra material was removed to avoid contamination. The crystal boule was oriented using a Multiwire Laboratories, model MWL110 Laue camera. Oriented single crystals were cut from the middle of the single crystal boule for magnetization and heat capacity measurements. Pieces were cut so the field could be applied along the [100], [110], and [111] directions on the same piece.

### Structure determination

Single crystal X-ray diffraction (SXRD) determination of the crystal structure of $NaCaNi_2F_7$ was performed on a Bruker diffractometer equipped with an Apex II detector and graphite monochromated Mo Kα radiation at 100 K. Data collection used the Bruker APEXII software package and subsequent reduction and cell refinement were performed using Bruker SAINT software[21]. The crystal structure was determined through the use of SHELXL-2013 software as implemented through the WinGX software suite[22,23]. The bulk material in the boule of $NaCaNi_2F_7$ was confirmed to have the pyrochlore structure by PXRD collected at 300 K using a Bruker D8 Eco with a LynxEye XE detector (Cu Kα). PXRD was collected on a single crystal ground in an argon-filled glove box and loaded into an airtight sample holder. The PXRD pattern was fit with the Thompson-Cox-Hastings pseudo Voight profile convoluted with axial divergence asymmetry through the FullProf software suite[24]. A unit cell of 10.3027 (2) Å was found for the bulk pieces of the $NaCaNi_2F_7$ by least squares fitting of the PXRD pattern. This is in good agreement with the single-crystal diffraction and the previous report[3].

Tables:

**Table 1: Single Crystal Data and Structural Refinement for NaCaNi$_2$F$_7$**

| | |
|---|---|
| Formula weight | 313.443 g/mol |
| Crystal System | Cubic |
| Space Group | $Fd\bar{3}m$ (227, origin 2) |
| Unit Cell | a=10.280(2) Å |
| Volume | 1086.4(7) Å$^3$ |
| Z | 8 |
| Radiation | Mo Kα |
| T | 100 K |
| Absorption Coefficient | 8.032 |
| F(000) | 1200 |
| Reflections collected/unique | 2929/87 R$_{int}$ = 0.0211 |
| Data/Parameters | 87/10 |
| Goodness-of-fit | 1.216 |
| Final R indices [I>2σ(I)] | R$_1$=0.0187, wR$_2$=0.0490 |
| Largest diff. peak and hole | 0.23 and -0.65 e A$^{-3}$ |

**Table 2: Atomic positions and anisotropic thermal displacement parameters for NaCaNi$_2$F$_7$**

| | Site | x | y | z | Occ. | U11 | U22 | U33 | U23 | U13 | U12 |
|---|---|---|---|---|---|---|---|---|---|---|---|
| Na | 16d | 0.5 | 0.5 | 0.5 | 0.5 | 0.0107(5) | 0.0107(5) | 0.0107(5) | -0.0029(3) | -0.0029(3) | -0.0029(3) |
| Ca | 16d | 0.5 | 0.5 | 0.5 | 0.5 | 0.0107(5) | 0.0107(5) | 0.0107(5) | -0.0029(3) | -0.0029(3) | -0.0029(3) |
| Ni | 16c | 0 | 0 | 0 | 1 | 0.0024(3) | 0.0024(3) | 0.0024(3) | -0.00011(13) | -0.00011(13) | -0.00011(13) |
| F(1) | 8b | 0.375 | 0.375 | 0.375 | 1 | 0.0052(8) | 0.0052(8) | 0.0052(8) | 0 | 0 | 0 |
| F(2) | 48f | 0.3303(2) | 0.125 | 0.125 | 1 | 0.0080(9) | 0.0072(6) | 0.0072(6) | 0.0039(8) | 0 | 0 |